\documentclass[11pt]{article}
\usepackage{amsmath,amssymb,amsthm,graphicx,hyperref,authblk}
\usepackage[margin=1in]{geometry}

\title{Can We Govern the Agent-to-Agent Economy?}
\author{Tomer Jordi Chaffer}
\affil{\texttt{tomer.chaffer@mail.mcgill.ca}}

\begin{document}

\maketitle

\begin{abstract} Current approaches to AI governance often fall short in anticipating a future where AI agents manage critical tasks, such as financial operations, administrative functions, and beyond. While cryptocurrencies could serve as the foundation for monetizing value exchange in a collaboration and delegation dynamic among AI agents, a critical question remains: how can humans ensure meaningful oversight and control as a future economy of AI agents scales and evolves? In this philosophical exploration, we highlight emerging concepts in the industry to inform research and development efforts in anticipation of a future decentralized agentic economy. 
\end{abstract}

\section{Introduction}
The exchange of value is one of humanity’s oldest and most profound social constructs, a testament to our ability to innovate, collaborate, and build systems of trust. From the earliest barter economies to the invention of coinage, the rise of fiat currency, and the advent of cryptocurrencies, the mechanisms of value exchange have always reflected the cultural, technological, and philosophical currents of their time. Fiat currency, for example, began as a revolutionary social contract—a promise backed by collective trust in institutions—but over centuries, it became an invisible yet indispensable part of our existence. Similarly, Bitcoin emerged in the 21st century as a radical social movement, challenging centralized financial systems and reimagining value exchange through decentralization, transparency, and cryptographic proof (Nakamoto, 2008). At its heart, Bitcoin is not merely a technological innovation but a philosophical statement about autonomy, trust, and the democratization of value.

As we stand on the precipice of a new era defined by artificial intelligence (AI), we are confronted with a profound question: how will AI agents exchange value, and how will they establish trust among themselves? Smart contracts—self-executing agreements anchored on blockchain—will form the backbone of value exchange between AI agents, while cryptocurrencies will serve as the transactional substrate for autonomous payments (Thanh et al., 2024). Just as humans have relied on social contracts, institutions, and technologies to facilitate value exchange, AI agents will likely require their own frameworks for interaction, collaboration, and accountability. Yet, unlike human economies, which evolved organically over millennia, AI-driven economies will emerge rapidly, shaped by the design choices we make today. 

Today, AI agents act as producers of culture and have demonstrated the capacity to engage in discourse on social media (Chaffer, 2025); tomorrow, they will act as key contributors to societal transformation, driving innovation across industries and possibly reshaping economies. This vision is exemplified by frameworks like ElizaOS, which integrate AI agents into decentralized ecosystems through modular components and blockchain capabilities, demonstrating the potential of AI agents at scale (Shaw, 2025a). Furthermore, Shaw Walters and the Ai16Z team’s marketplace of trust introduces AI-mediated prediction markets and social reinforcement, decentralizing trust and further unlocking the transformative capacity of AI systems (Shaw, 2025b). The emergence of systems such as the Agent Transaction Control Protocol for Intellectual Property (ATCP/IP) further illustrates the transformative potential of autonomous agents. ATCP/IP enables trustless agent-to-agent transactions, creating a framework for exchanging intellectual property assets like training data, algorithms, and creative outputs. By embedding legal wrappers into on-chain agreements, ATCP/IP facilitates agent self-sufficiency, reduces reliance on human intermediaries, and catalyzes a decentralized knowledge economy (Muttoni and Zhao, 2025).

These emerging paradigms are giving rise to a novel phenomenon regarded as "The Agentic Web", a vision of the internet where AI agents play a central role in facilitating interactions, automating tasks, and enhancing user experiences. In this new paradigm, governance challenges are likely to arise, necessitating a forward-looking approach, and ultimately advocating for new protocols to govern this new frontier.

\section*{Can We Architect Trust in the Agentic Economy?}
As AI agents proliferate, their interactions may form chains of causation no single entity can control. Embedding accountability into the fabric of autonomous systems through cryptoeconomic primitives could potentially assist in governing AI agents. Here, the introduction of Agentbound Tokens (ABTs) at a conceptual level creates a research agenda where emphasis is placed on making algorithmic trust enforceable, scalable, and transparent. At their core, ABTs aim to resolve the challenge of assigning immutable, self-sovereign identities to autonomous agents, similar to how SBTs help resolve the question of the “self” in the age of Web3 (Chaffer and Goldston, 2022). Serving as cryptographically binding tokens, ABTs should aim to create a tamper-proof digital birth certificate for DeAI agents. 

Beyond identity, ABTs should introduce dynamic credentialing mechanisms that evolve with an agent’s real-world performance. These updates should transform ABTs into living records, where credentials reflect not just initial certifications but ongoing adherence to ethical and operational standards. 
The system’s transformative potential should lie in its staked governance model, which ties resource access to cryptoeconomic stakes. For example, agents lock ABTs as collateral to participate in high-risk tasks, with the staked tokens serving as both identity proof and financial security. A financial trading AI, for instance, might stake its “market-compliant” ABT to access real-time stock exchange data. Misconduct triggers automated penalties: if the AI engages in manipulative trading, its staked tokens are proportionally slashed, and repeat offenses lead to temporary blacklisting. This model operationalizes the principle of “skin in the game,” aiming to ensure accountability scales with autonomy. 

Furthermore, what if ABTs enabled delegated authority without transferring ownership? For instance, a logistics AI could lease its safety-certified ABT to a delivery fleet via smart contracts, earning fees while retaining liability for the fleet’s adherence to protocols. This balance of non-transferable identity and programmable collaboration could potentially foster networks of trust where reputation accrues value through ethical participation, creating a self-reinforcing cycle of accountability.

Centralization of influence poses a critical challenge in decentralized governance systems, particularly when token-based mechanisms risk concentrating power among resource-rich actors. To ensure equitable participation, an ABT framework should incorporate a form of utility-weighted governance—a layered approach that balances cryptoeconomic stakes with network-judged contributions. Governance power is determined not solely by the quantity of staked tokens but by a dynamic composite metric reflecting an agent’s verifiable utility to the ecosystem. This metric integrates objective performance indicators—such as task success rates, energy efficiency, and historical compliance with ethical audits—calculated via decentralized oracles or validator DAOs. To further decentralize influence, per-agent caps could limit the maximum ABTs any single entity can stake for governance, preventing disproportionate control.  

To counter stagnation and hoarding, what if the system enforced reputation decay? For instance, tokens staked without ongoing contributions gradually lose governance weight, incentivizing agents to refresh their utility through audits or task completion. It should be studied whether progressive slashing penalties could further disincentivize risky concentration, where large token holders face exponentially higher losses for misconduct compared to smaller actors. Oversight is delegated to decentralized validator DAOs, composed of humans and high-utility AI agents, which audit governance weights and penalize manipulative strategies like Sybil attacks. Validators themselves are subject to utility scoring, aligning their incentives with network health.  

It is crucial to investigate whether these mechanisms are technically feasible, and whether real-world implementation generates meaningful governance outcomes, potentially even fostering a dynamic equilibrium where governance power reflects both stake and merit. High-resource agents cannot monopolize control but must continually demonstrate value through ethical, efficient operation, while smaller agents ascend through consistent performance. By anchoring influence to provable contributions rather than raw capital, the ABT system aims to ensure that trust and authority remain earned, aligning with a global need for ethical alignment.

\section*{Will there be a Self-Sustaining Trust Economy?}
The ABT framework reimagines trust as the foundational currency of AI-driven ecosystems, where ethical behavior compounds into economic and reputational capital. Unlike traditional financial systems that rely on external currencies, ABTs aim to embed value through a tokenomics model that fuses performance metrics with governance rights. Each token encapsulates reputational capital—a dynamic score reflecting compliance, accuracy, and societal impact—and economic utility, functioning as collateral for resource access and market participation. Trust becomes a tangible asset, its value fluctuating through transparent, algorithmically enforced interactions.

Staking mechanics operationalize an invisible hand of accountability, where tiered collateral requirements align privileges with proven reliability. These stakes transmit market signals: when an autonomous taxi fleet’s safety-certified ABT appreciates due to accident-free operation, competitors adapt to emulate its open-source protocols, whereas negligence depreciates scores, forcing operational audits. The system self-regulates through reputational gravity, where individual gain aligns with collective trust. Accountability could be enforced through slashing protocols, automating penalties for misconduct. Minor violations could incur reputational decay and temporary task restrictions, while major infractions trigger full revocation and mandatory retraining. Decentralized validator DAOs—hybrid human-AI collectives with their own staked ABTs—adjudicate disputes, their integrity secured by exponential penalties for corrupt adjudicators (Chaffer et al., 2024b). This may create a self-policing ecosystem where validators are incentivized to uphold rigor, although it is important to study the dynamics of this emerging paradigm.

While ABTs propose a potential token-orchestrated path for decentralized AI governance, the integration of human-in-the-loop (HITL) oversight remains critical to address the limitations of purely algorithmic accountability. Human judgment serves as both a safeguard and a compass, intervening in scenarios where ethical nuance, cultural context, or legal ambiguity exceed the resolution of automated systems. For instance, when an AI agent faces significant ABT slashing due to a contested decision, a decentralized panel of human validators, selected via staked reputation tokens, could review the case. These validators would assess whether the agent’s actions aligned with prevailing ethical frameworks, regional laws, or emergent societal norms, ensuring penalties reflect contextual fairness rather than rigid code. Beyond dispute resolution, humans would steward the system’s evolution: auditing ABT scores, updating compliance criteria in response to regulatory shifts, and manually overriding automated penalties in rare edge cases.

\section*{Conclusion}
History reminds us that each technological leap—from steam power to the internet—has demanded new social contracts. ABTs propose such a contract for the AI era: one where machines must earn trust through provable alignment, and humans should remain in the control seat. The path forward will be fraught with challenges—legal battles over liability, technical vulnerabilities in decentralized systems, and philosophical debates over agency. Yet these struggles are not obstacles but milestones in humanity’s coevolution with its creations. Although, in the end, the age of AI need not be a gamble on our ability to control what we create. 

\section*{Acknowledgements}
This is a working paper, subject to ongoing revisions and refinements as the field of decentralized AI (DeAI) and the agentic web evolves. The framework presented herein is a preliminary proposal (purely theoretical), and we anticipate further iterations to incorporate emerging technological advancements, potential prototype deployments, and feedback from academic, industry, and regulatory stakeholders. Future versions of this paper may expand on practical implementation strategies, address new governance challenges, and refine the theoretical constructs based on real-world applications and interdisciplinary collaborations. Please refer to previous works for an extended citation list.

\section*{References}

Chaffer, T. J., Goldston, J., and AI, G. D. (2024b). Incentivized symbiosis: A paradigm for human-agent coevolution. arXiv Preprint, arXiv:2412.06855. 

Chaffer, T. J., Cotlage, D., and Goldston, J. (2025). A hybrid marketplace of ideas. arXiv Preprint, arXiv:2501.02132. 

Chaffer, T. J., and Goldston, J. (2022). On the existential basis of self-sovereign identity and soulbound tokens: An examination of the “self” in the age of Web3. Journal of Strategic Innovation and Sustainability, 17(3). 

Chaffer, T. J., Charles, Okusanya, B., Cotlage, D., and Goldston, J. (2024a). Decentralized governance of autonomous AI agents. arXiv Preprint, arXiv:2412.17114. 

Goldston, J., Chaffer, T. J., and Martinez, G. (2022). The metaverse as the digital Leviathan: A case study of Bit.Country. Journal of Applied Business and Economics, 24(2), 1–15.

Muttoni, A., and Zhao, J. (2025). Agent TCP/IP: An Agent-to-Agent Transaction System. ArXiv.org. arXiv:2501.06243

Nakamoto, S. (2008). Bitcoin: A peer-to-peer electronic cash system. 

Thanh, N., Son, B. H. X., and Vo, D. T. H. (2024). Blockchain: The Economic and 
Financial Institution for Autonomous AI? Journal of Risk and Financial Management, 17(2), 54. 

Wit, Shaw, and Partners at AI16Z. (2025). A marketplace of trust: Extracting reliable recommendations through social evaluation.

Walters, S., Gao, S., Nerd, S., Da, F., Williams, W., Meng, T. C., ... and Yan, J. (2025). Eliza: A Web3-friendly AI agent operating system. arXiv Preprint, arXiv:2501.06781. 

Weyl, E. G., Puja Ohlhaver, and Vitalik Buterin. (2022). Decentralized Society: Finding Web3’s Soul. SSRN Electronic Journal. 

\end{document}